# Modification of Symmetric Cryptography with Combining Affine Chiper and Caesar Chiper which Dynamic Nature in Matrix of Chiper Transposition by Applying Flow Pattern in the Planting Rice

Dewi Sartika Ginting[*,1], Kristin Sitompul[1], Jasael Simanulang[1], Rahmat Widia Sembiring[2], Muhammad Zarlis[3]

[1]*Faculty of Computer and Information Technology, Universitas Sumatera Utara, Medan, 20155, Indonesia*

[2]*Politeknik Negeri Medan, Medan, 20155, Indonesia*

[3]*University of North Sumatera, Medan, 20155, Indonesia*

*Email: dewigintingdg90@gmail.com, jasaelsimanullang@yahoo.co.id, rahmatws@polmed.ac.id, m.zarlis@usu.ac.id*

A B S T R A C T

*Classical cryptography is a way of disguising the news done by the people when there was no computer. The goal is to protect information by way of encoding. This paper describes a modification of classical algorithms to make cryptanalis difficult to steal undisclosed messages. There are three types of classical algorithms that are combined affine chiper, Caesar chiper and chiper transposition. Where for chiperteks affine chiper and Caesar chiper can be looped as much as the initial key, because the result can be varied as much as key value, then affine chiper and Caesar chiper in this case is dynamic. Then the results of the affine and Caesar will be combined in the transposition chiper matrix by applying the pattern of rice cultivation path and for chipertext retrieval by finally applying the pattern of rice planting path. And the final digit of the digit shown in the form of binary digits so that 5 characters can be changed to 80 digit bits are scrambled. Thus the cryptanalyst will be more difficult and takes a very long time to hack information that has been kept secret.*

## 1. Introduction

A security issue is one of the most important aspects of an Advancements in the field of computer technology allows thousands of people and computers around the world connected in a virtual world known as *cyberspace* or Internet. But unfortunately, technological advances are always followed by a downside to the technology itself. One is the susceptibility of data security, giving rise to the challenges and demands the availability of a data security system is as sophisticated as the technology advances of the computer itself. This is the background of the development of a data security system to protect data transmitted through a communications network [1]. There are several ways to do security data through one channel, one of which is cryptography. In cryptography, data transmitted via the network will be disguised in a way that even if the data can be read then it cannot be understood by unauthorized parties. Data to be transmitted and not experienced isitilah encoding known as plaintext, and after camouflaged with a way of encoding, then this will turn plaintext into ciphertext. The functions that are fundamental in cryptography is encryption and decryption [2].

Many cryptographic techniques are implemented to safeguard information, but the present condition is much too way or the work done by cryptanalysis to break it. Though an important thing in the delivery of the message is to maintain the security of the information that are not easily known or manipulated by other parties. One solution that can be done is to modify the cryptography solved or create a new cryptography so that it can be an alternative for securing messages [3].

[*]Corresponding Author: Dewi Sartika Ginting, Faculty of Computer and Information Technology, Universitas Sumatera Utara, Medan, 20155, Indonesia
Email: dewigintingdg90@gmail.com

This study design a symmetric key cryptography by using 3 classical cryptographic algorithm, of which two are affine and Caesar chiperteksnya results will be dynamic (changing). And the matrix transposition cipher is a medium to combine chiperteks affine and Caesar on the technique of laying of the bits in the matrix using the rice planting furrow pattern and sampling technique of bits to be chiperteks end using the groove pattern-making transplanting rice. And in this study chiperteks be displayed in the form of binary numbers are 1 and 0. This research could be useful as a modification that serve as an alternative cryptographic security of information, so that the cryptanalyst would have difficulty or even require a very long time in the hacking of information [4-10].

## 2. Theoretical Basic

Encryption is the process of converting a plain text to be a message in the ciphertext.

C = E (M)

Where:
M = original message
E = encryption process
C = message in code (for brevity called the password), while

Decryption is the process of changing the password message in one language into the original message back.

M = D (C)

Where:
D = the decryption process

Generally, apart from using certain functions in the encryption and decryption, it is often a function was given an additional parameter called the key terms. For example, the original text: "experience is the best teacher". Once encrypted password algorithm with key xyz and PQR into cipher text: V583ehao8 @ $%.

### 2.1. Algorithm of Affine Chiper

Affine cipher on the method of affine is an extension of the method Caesar Cipher, which divert the plaintext with a value and add it with a shift of P produce ciphertext C is expressed by the function of congruent:

$$C \equiv m P + b \pmod{n}$$

In which n is alphabet size, m is an integer which must be relatively prime to n (if not relatively prime, then the decryption can not be done) and b is the number of shifts (Caesar cipher is the specialty of affine cipher with m = 1). To carry out the description, equation (4) herus solved to obtain a solution P. congruence exists only if inver m (mod n), expressed in m $^{-1}$. Jikam $^{-1}$ exists then decryption is done by the following equation:

$$P \equiv m^{-1}(C - b) \pmod{n}$$

### 2.2. Algorithm of Caesar Chiper

Cryptographic one of the oldest and simplest is Caesar cryptography. Historically, this is how Julius Caesar send important letters to the governors. Caesar cryptographic formula, generally can be written as follows:

C = E (P) = (P + k) mod 26
P = D (C) = (C k) mod26, where
P is the plaintext,
C is the ciphertext,
K is shifting letter corresponding to the desired key.

### 2.3. Algorithm of Chiper Transposition

Cryptographic in column (column cipher), plaintext letters are arranged in groups consisting of several letters. Then the letters in this group write a column by column, the order of columns can be fickle. Cryptography column is one example of cryptographic methods of transposition. Example Cryptography Column:

The phrase 'FATHER HAS ARRIVED YESTERDAY AFTERNOON', if arranged in columns 7 letter, it will be the following columns:

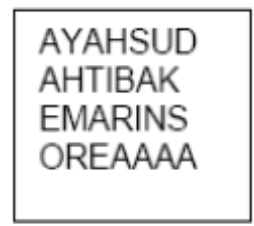

**Figure 1**. Structure of Character Chiper Transposition

To complete the last column that contains 7 letters, then the rest is filled with the letters' A 'can be any letter or as a complementary letter. Tesebut sentence after 7 columns encrypted with a key sequence of letters and 6,725,431, the result of the encryption:

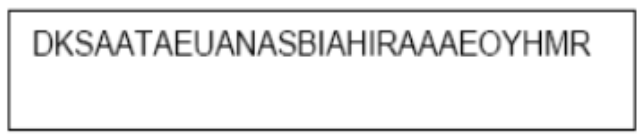

**Figure 2**. Column Encryption of Chiper Transposition

## 3. Research Methods

### 3.1. Research Design

In the development of classical cryptography is included 3 classic algorithm that is affine cipher, Caesar cipher and a transposition cipher. Wherein the plaintext to be executed by using affine cipher and the Caesar cipher in advance separately, and in the encryption process is dynamic, ie chiperteks generated by affine and caesarean may change according to the wishes, of which a maximum of change as the keys that are used at the beginning of the process affine and Caesar.

Changes that occur chiperteks generated by looping the same process. Below we can see the design of stages of research:

And to process returns to the plaintext message from chiperteks called process descriptions, can be seen in the design below:

*3.2. 3.2. Limitations of Problem*

To not extend the scope of the discussion of the given limitations in this study, namely;
1. The process of encryption and decryption is done on the text.
2. Key b at the same affine with key k at Caesar specified by the author of the message and should be a number, whereas m in the affine key is a prime number.

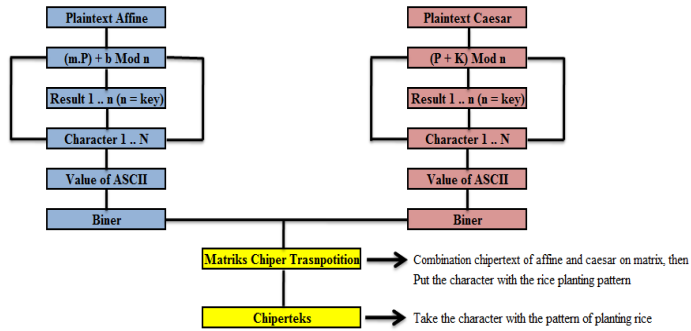

**Figure 3.** Research Design Process Encryption

3. Affine cipher algorithm and a Caesar cipher generates a dynamic chiperteks, where the results are changed just as much as the initial key specified by the author of the message.
4. Recurrence happened not more than the limit, and the limit on the initial key = used affine key (b) and Caesar (key k).
5. Laying chiperteks affine and Caesar on Marik transposition cipher using rice cropping patterns and making the character to chiperteksnya is to use pattern-making transplanting rice.

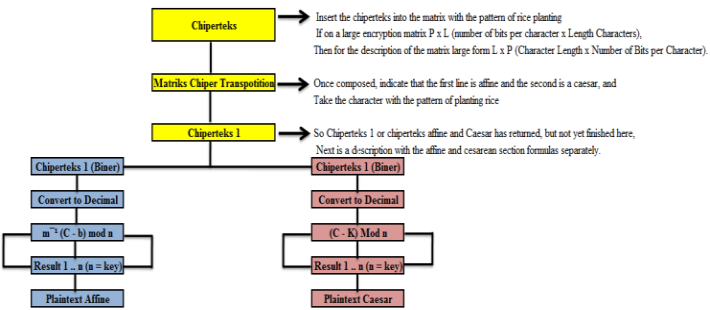

**Figure 4.** Process Research Design Description

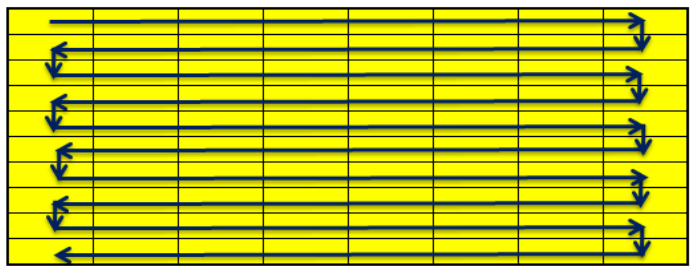

**Figure 5.** Pattern of Rice Planting

## 4. Results and Discussions

This study to design the development of classical cryptography involving three classic algorithms are affine cipher, Caesar cipher and a transposition cipher, where the encryption result of the affine cipher and the Caesar cipher is dynamic. And matrices on transposition cipher contain binary blend chiperteks of affine cipher and the Caesar cipher.

*4.1. Rice Planting Pattern Flow*

Method rice planting is usually done in a long horizontal continuous with mapped fields. This study used the same way as the planting of rice, using a matrix transposition ciphe

*4.2. Flow Pattern Making Rice Planting*

Rice planting take groove pattern is used as a pattern for the process description, which is to get back the encrypted plaintext before the rice planting groove pattern on the matrix transposition cipher.

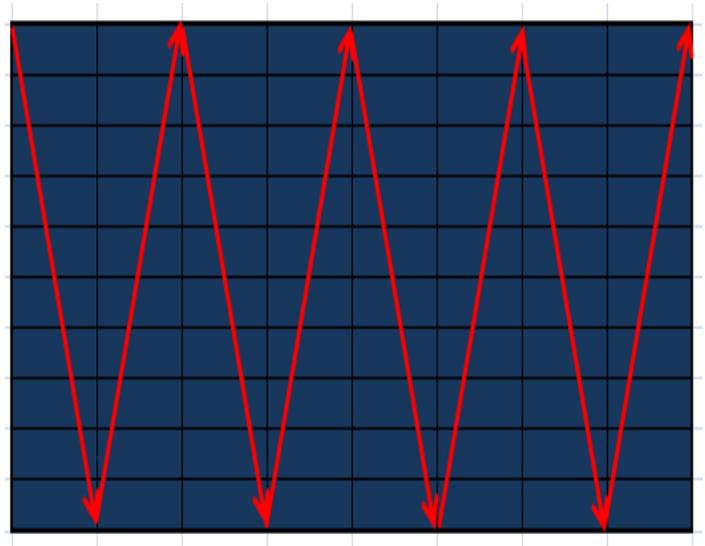

**Figure 6.** Flow Patterns Rice Planting Decision

*4.3. Flowchart of Encryption*

In cryptography this modification, first in plaintext encryption process we execute with affine cipher and the Caesar cipher separately. Where the results of the affine cipher encryption and later Caesar cipher is a dynamic, namely chiperteks generated by each algorithm can be fickle. It is very helpful to add to the confusion of the cryptanalyst. And to create dynamic results, the authors do looping on the affine cipher or Caesar cipher formula denagn respectively. But in this study, we limit the repetition can be done, namely a maximum of keys are initialized at the beginning. For example, key in the beginning is 15, then looping may occur up to 15 times.

**a. Affine Chipper Encryption**

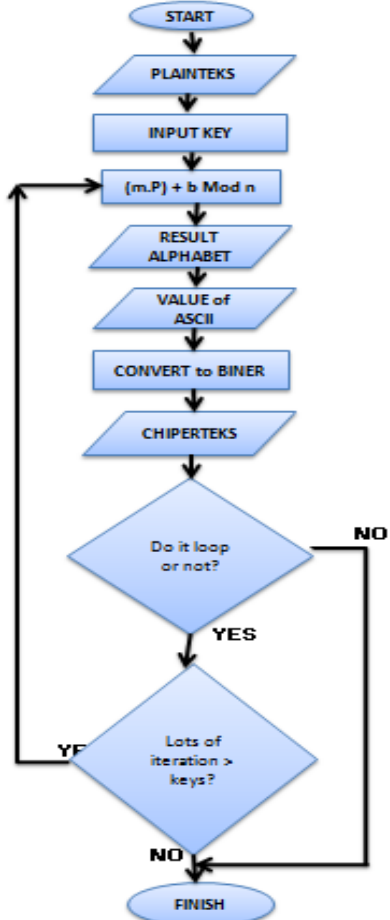

**Figure 7.** Flowchart of Affine Chiper

**b. Caesar Chiper Encryption**

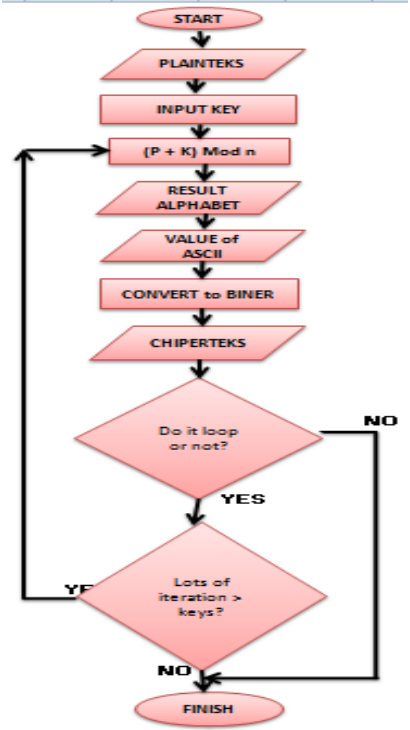

**Figure 8.** Flowchart of Caesar Chiper

**c. Encryption of Chiper Transposition**

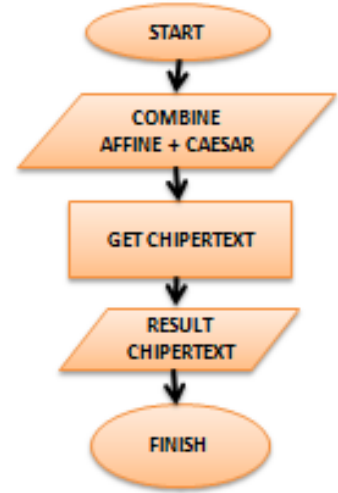

**Figure 9.** Flowchart encryption of Chiper Trasnposition

In transposition ciphers will be included chiperteks of affine cipher and the Caesar cipher to the matrix. Chiperteks of affine and Caesar in the form of binary digits, and the digits of the binary is placed into the matrix to follow the pattern of grooves cropping with the provisions of the first line starts from the affine cipher followed by a Caesar cipher and so on so forth until all the binary digits chiperteks on affine and Caesar composed both in the matrix transposition cipher.

*4.4. Flowchart of Description*

For the description of the process of restoring the messages that have been encrypted, or in other words to restore the plaintext of chiperteks, starting from a transposition cipher.

a. Description of Transposition Chiper

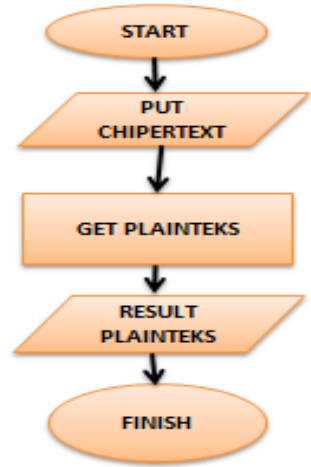

**Figure 10**. Flowchart Description of Chiper Trasnposisi

After receiving the plaintext of a transposition cipher, the plaintext is chiperteks results affine and Caesar cipher encryption. It is still necessary for the next stage of the description that the message be the same again. And the next process is mendeksripsikan the description of a transposition cipher using a formula descriptions affine cipher and the Caesar cipher.

**b. Affine Chiper Description**

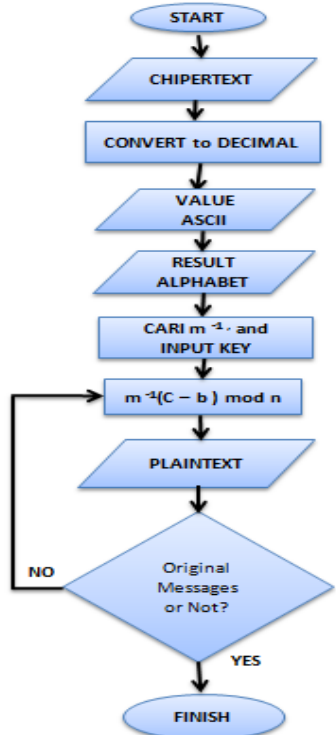

**Figure 11.** Flowchart of Affine Chiper

**c. Chipper Description**

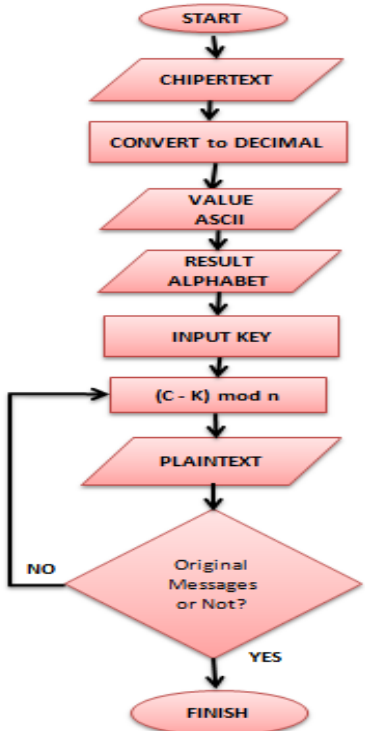

**Figure 12**. Flowchart of Caesar Chiper

*4.5. Discussion of combined Affine chiper and Caesar chiper in the matrix of Chiper Transposition*

For processes in detail, if we presuppose the plaintext is Y and chiperteks is C were converted into bits, then:

Y = {Y1, Y2, Y3, Y4, Y5, ..., Yn}

Y1 = {C1, C2, C3, ..., C8}
Y2 = {C9, C10, C11, ..., C16}
Y3 = { C17, C18, C19, ..., C24}… and so on up to Yn.

To process the cipher transposition, then we take chiperteks on affine cipher and the Caesar cipher and place in the matrix with the provisions of the algorithm is:

The first line is the first character or that has been converted into 8-digit bit chiperteks of affine, which started with a pattern of grooves rice planting ie from left to right on the first row (odd) and so on.

Then for the second row, followed by the first character of the Caesar cipher chiperteks or 8-digit bit after conversion is done according to the rules also cropping the groove pattern from left to right of the second row (even) and so on.

**So can we symbolize results occur:**

| C1.A | C2.A | C3.A | C4.A | C5.A | C6.A | C7.A | C8.A |
|---|---|---|---|---|---|---|---|
| C8.B | C7.B | C6.B | C5.B | C4.B | C3.B | C2.B | C1.B |
| C9.A | C10.A | C11.A | C12.A | C13.A | C14.A | C15.A | C16.A |
| C16.B | C15.B | C14.B | C13.B | C12.B | C11.B | C10.B | C9.B |
| C17.A | C18.A | C19.A | C20.A | C21.A | C22.A | C23.A | C24.A |
| C24.B | C23.B | C22.B | C21.B | C20.B | C19.B | C18.B | C17.B |
| C25.A | C26.A | C27.A | C28.A | C29.A | C30.A | C31.A | C32.A |
| C32.B | C31.B | C30.B | C29.B | C28.B | C27.B | C26.B | C25.B |
| C33.A | C34.A | C35.A | C36.A | C37.A | C38.A | C39.A | C40.A |
| C40.B | C39.B | C38.B | C37.B | C36.B | C35.B | C34.B | C33.B |

Description : A symbol is Affine Chiper, and
B Symbol is Caesar Chiper.

**Figure 13.** Matrix Encryption of Chiper Transposition (Put plaintext)

See matrix above, the number of bits is 8 x 10 = 80 bits, which consists of 40 bits and 40 bits affine Caesar. And the last step in the process of encryption to obtain chiperteks the above matrix is to apply the rice cropping making workflow, from left column top to bottom and so on.

The matrix can be seen in the figure below:

| C1.A | C2.A | C3.A | C4.A | C5.A | C6.A | C7.A | C8.A |
|---|---|---|---|---|---|---|---|
| C8.B | C7.B | C6.B | C5.B | C4.B | C3.B | C2.B | C1.B |
| C9.A | C10.A | C11.A | C12.A | C13.A | C14.A | C15.A | C16.A |
| C16.B | C15.B | C14.B | C13.B | C12.B | C11.B | C10.B | C9.B |
| C17.A | C18.A | C19.A | C20.A | C21.A | C22.A | C23.A | C24.A |
| C24.B | C23.B | C22.B | C21.B | C20.B | C19.B | C18.B | C17.B |
| C25.A | C26.A | C27.A | C28.A | C29.A | C30.A | C31.A | C32.A |
| C32.B | C31.B | C30.B | C29.B | C28.B | C27.B | C26.B | C25.B |
| C33.A | C34.A | C35.A | C36.A | C37.A | C38.A | C39.A | C40.A |
| C40.B | C39.B | C38.B | C37.B | C36.B | C35.B | C34.B | C33.B |

**Figure 14.** Matrix Encryption of Chiper Transposition (Decision Chiperteks)

From the matrix above, we can take chiperteks by following the rice planting groove pattern as above, then chipertek is:

C ={C1A, C8B, C9A, C16B, C17A, C24B, C25A, C32B, C33A, C40B, C39B, C34A, C31B, C26A, C23B, C18A, C15B, C10A , C7B, C2A, ... ... ..., C33B, C40A, C25B, C32A, C27B, C24A, C9B, C16A, C1B, C8A}

Chiperteks that have been obtained in the above then we descriptions to get the plaintext back. This is what will be done recipient messages or information. The algorithm description on transposition cipher chiperteks above is:

1. Put chiperteks transposition cipher encryption on the results into the matrix, where the matrix is a comprehensive description width x length (number of characters x number of bits per character) or the inverse of the encryption process. Peletakannya same way with encryption that is by adopting a rice planting furrow.

**Figure 15**. Matrix Description of Chiper Transposition (Put Chiperteks)

2. After putting chiperteks into the matrix, then the next step is to take the initial plaintext or chiperteks of affine and Caesar by following the flow of the rice-planting decision.

**Figure 16.** Matrix Description of Chiper Transposition (Plaintext Intake / Chiperteks1)

3. And can be seen in the matrix with the pattern of grooves taking over the rice planting, that plaintext or chiperteks 1 which means chiperteks of affine and Caesar already apparent. If we write the tracks arrow then these bits are arranged well and return to the previous message. And it seems clear also that the symbol is a bit affine cipher A and B are Caesar cipher. The results of the description if written is:

P = {P1A, P2A, P3A, P4A, P5A, P6A, P7A, P8A, C1B, C2B, C3B, C4b, C5b, C6b, C7B, C8B, C9A, C10A, C11A, C12A, C13A, C14A, C15A, C16A ... ... ..., C33A, C34A, C35A, C36A, C37A, C38A, C39A, C40A, C33B, C34B, C35B, C36B, C37B, C38B, C39B, C40B}

*4.6. Test Result*

There are some that can can of this algorithm, the invention includes some strengths and weaknesses in this algorithm. Namely:

a. Advantages :

1. The results of the dynamic at chiperteks early stage in the process of the affine cipher and the Caesar cipher, becomes a strength because the results can vary and this will make the cryptanalyst becomes more difficult to hack.

2. Chiperteks end that is sent to the message recipient is the number of binary digits, so if 5 characters or one word shall we send, will generate 80 digit bit. And even this 80-digit randomly arranged by rice cropping pattern groove, so that in testing it becomes an advantage in this algorithm.

b. Weaknesses :

However, there are things that still looks weak in the algorithm, namely a recurrence that may be performed in affine and Caesar does not exceed the value of the key early, it can be said that the larger the key will be the better, but if the keys are initialized at the beginning of the small, then the less recurrence happens that a hacker does not require a long time to try to hijack the message.

5. **Conclusions**

a. Chiperteks dynamic process and Caesar affine cipher is also an important point on the modification of this algorithm, as plaintext consists of one message alone can generate an arbitrary chiperteks. And this technique can also trickthe cryptanalyst.

b. The use of algorithms rice planting groove pattern on the matrix transposition cipher can prove that this technique generates a symmetric cryptographic methodology stronger.

c. The combination of several methods on classical cryptography such as affine cipher, Caesar cipher and a transposition cipher increase the level of difficulty of this cryptography. The encryption process is layered with berebeda method will make hackers need at a very long time to steal the information or message.

d. Chiperteks display in the form of binary digits that doubled the number of bits in each character, will help add to the confusion cryptanalyst for one word alone can produce tens or hundreds of binary bits consist of 1 and 0.